\documentclass[aps,pre,twocolumn,showpacs]{revtex4-1}
\usepackage{epsfig}

\begin{document}

\title{Determining efficient temperature sets for the simulated tempering 
method} 

\author{A. Valentim, M. G. E. da Luz}
\affiliation{Departamento de F\'{\i}sica,
Universidade Federal do Paran\'a,
CP 19044, 81531-980 Curitiba-PR, Brazil}
\author{Carlos E. Fiore}
\affiliation{Instituto de F\'{\i}sica,
Universidade de S\~{a}o Paulo, \\
Caixa Postal 66318\\
05315-970 S\~{a}o Paulo, S\~{a}o Paulo, Brazil}
\date{\today}

\begin{abstract}
In statistical physics, the efficiency of tempering approaches strongly 
depends on ingredients such as the number of replicas $R$, reliable 
determination of weight factors and the set of used temperatures, 
${\mathcal T}_R = \{T_1, T_2, \ldots, T_R\}$.
For the simulated tempering (SP) in particular -- useful due to its 
generality and conceptual simplicity -- the latter aspect (closely related 
to the actual $R$) may be a key issue in problems displaying metastability 
and trapping in certain regions of the phase space.
To determine ${\mathcal T}_R$'s leading to accurate thermodynamics 
estimates and still trying to minimize the simulation computational time, 
here it is considered a fixed exchange frequency scheme for the ST.
From the temperature of interest $T_1$, successive $T$'s are chosen so that 
the exchange frequency between any adjacent pair $T_r$ and $T_{r+1}$ has a 
same value $f$.
By varying the $f$'s and analyzing the ${\mathcal T}_R$'s through relatively 
inexpensive tests (e.g., time decay toward the steady regime), an optimal 
situation in which the simulations visit much faster and more uniformly 
the relevant portions of the phase space is determined.
As illustrations, the proposal is applied to three lattice models, BEG, 
Bell-Lavis, and Potts, in the hard case of extreme first-order phase 
transitions, always giving very good results, even for $R=3$.
Also, comparisons with other protocols (constant entropy and arithmetic 
progression) to choose the set  ${\mathcal T}_R$ are undertaken.
The fixed exchange frequency method is found to be consistently superior, 
specially for small $R$'s.
Finally, distinct instances where the prescription could be helpful 
(in second-order transitions and for the parallel tempering approach)  
are briefly discussed.
\end{abstract}


\maketitle

\section{Introduction}

Keystone in the study of statistical physics problems, numerical methods 
are generally expected to fulfil two requirements: (i) first (and surely the 
most important), to yield precise estimates for the thermodynamical quantities
analyzed; (ii) second, to be as simple and as fast as possible in their 
implementations.

Nevertheless, often the mentioned two requisites strike out in opposite 
directions.
Indeed, consider, e.g., systems in the regime of phase transitions whose 
distinct regions of the phase space are separated by large free-energy 
barriers.
It is a common situation not only for complex problems like spin glasses, 
protein folding, and biomolecules conformation \cite{spinglass,proteins,
pande22,allen,frenkel}, 
but also in lattice gas models displaying first-order phase transitions 
\cite{binder,landau}.
In all such examples there may be the occurrence of metastability 
\cite{binder,landau}.
Thus, when simulated, these systems can get trapped into local minima. 
Ways to circumvent this technical difficulty should demand more sophisticated 
evolution dynamics procedures and longer computational times.

Different proposals like, (a) cluster \cite{sw}, (b) multicanonical 
\cite{berg}, (c) Wang-Landau \cite{wang}, and (d) tempering 
\cite{nemoto,parisi}, among others, are relevant algorithms trying to 
maintain a good balance between features (i) and (ii) above.
In special, (d) above relies on the straightforward idea of ``heating up'' 
the system to higher temperatures, so to help it to cross the barriers at 
low temperatures. 
Moreover, tempering methods have attracted large interest due to their
generality with a broad applicability \cite{earl}.

There are two major formulations for the tempering approach, namely, 
parallel (PT) \cite{nemoto} and simulated (ST) \cite{parisi}, where 
always the start point is to choose a set of $R$ distinct temperatures 
(with  $T_r < T_{r+1}$, $r=1,\ldots,R-1$), $\mathcal{T}_R = 
\{T_1, T_2, \ldots, T_R\}$, in which $T_1$ is the one of interest.
In the PT, configurations from the distinct $R$ replicas (running in 
parallel) at the different $T$'s are exchanged.
For the ST, a single realization undergoes many temperature changes 
(among the $T$'s in $\mathcal{T}_R$).
Thus, the temperature itself is a dynamical variable.

Each tempering implementation presents its own characteristics and 
advantages, as recently discussed in details in \cite{fiore2} (see also 
the Refs. therein). 
In particular, although the ST has a higher probability than the PT to 
exchange temperature \cite{pande,park2,pande2,ma,fiore2}, it displays a 
less frequent tunneling between coexisting phases \cite{fiore2}.
Hence, the ST requires large computational time for generating uncorrelated 
configuration with a slower convergence to the steady equilibrium. 
On the other hand, for proper estimates (at least at first-order phase
transition regimes) the PT needs non-adjacent switch of temperatures, 
making the procedure a bit more involving -- an implementation not 
necessary for the ST.

Furthermore, for the hard to treat case of strong discontinuous 
phase transitions,
promising extensions for tempering methods have been proposed.
In particular, the PT combined with modified ensembles (as multiple 
Gaussians \cite{neuhaus-hager,neuhaus-magiera}) comprise the so called 
generalized replica exchange approaches 
\cite{kim-keyes-straub,kim-straub-keyes}.
They have been applied with great success to problems like solid-liquid 
\cite{lu1} and vapor-liquid transitions \cite{lu2}.
Also, enhancement for the usual ST are possible.
Examples are:
(a) to consider for it modified distributions \cite{kim-straub}, leading 
to very good results for both lattice (e.g., Potts and Ising) and continuum 
(Lennard-Jones clusters) models; and (b) besides $T$ to assume another 
dynamical variable, as the external field \cite{nagai-okamoto}, quite
helpful in dealing with crossovers in 2D Ising systems.

Thus, it would be desirable to improve the efficiency of the ST still
preserving its positive aspects, notably the procedure simplicity. 
As a hint to do so, the previous comments indicate that a central point 
in the ST method is less the probability of a single attempt to exchange
temperatures $T_{r} \rightarrow T_{r+n}$ (with $n=1$) and more the overall
frequency in which the different system phases are visited.
Therefore, one should try to optimize the set $\mathcal{T}_R$ as a whole, 
investigating how the combination of the different transitions would speed 
up the convergence to the steady state (by a more uniform sampling of the 
microscopic configurations).

For the ST we then propose here a rather direct protocol to select 
$\mathcal{T}_R$ by means of a fixed exchange frequency (FEF) prescription.
Given $R$, it consists in determining the $T_r$'s such that the exchange 
frequency between any pair of adjacent temperatures is $f$.
From simple preliminary tests we verify if the obtained set leads
to an appropriate tunneling between coexisting phases.
If this is not the case, another value of $f$ is chosen, a new  
$\mathcal{T}_R$ is calculated, and the tests repeated.
With relatively low computational effort (see next Section), we end up with 
a very efficient $\mathcal{T}_R$ for the full simulations.
Through examples, we furthermore show that this optimal $\mathcal{T}_R$
works well for other values of the considered parameters and not only 
for the specific values employed in the set derivation.
The same $\mathcal{T}_R$ also can be used in the vicinity of the original
parameters values as well as for other system sizes.
Hence, in many applications $\mathcal{T}_R$ needs to be determined just 
once.
We compare the FEF with other schemes to select the $T_r$'s.
We find that the present is not only superior to more simple recipes 
(like arithmetic progressions (AP)) but also to more physically oriented
selection methods (like the constant entropy (CE) \cite{sabo,fiorejcp}). 
We finally confirm a somehow expected result (but not fully investigated 
in the literature) that the exact distribution of temperatures in 
$\mathcal{T}_R$ becomes less relevant as $R$ increases.

As illustrations, we address three distinct systems, so exploiting
a relatively larger variety of first-order phase transitions features. 
One is the Potts model, an ideal case test.
For large $q$'s, it presents strong discontinuous transitions (the 
regime we shall focus), whose temperatures are exactly known.
The others are the BEG and BL models, likewise interesting not only 
by displaying more complex phase diagrams than the Potts (e.g., 
having phases with distinct structural properties), but also for 
already being extensively analyzed through the PT and ST approaches 
\cite{fiore2,fiorejcp,fiore3}.
Thus, all them are nice examples to check for the reliability of the
proposed protocol.

The work is organized as the following.
In Sec. II we review the ST approach and how to characterize 
first-order phase transitions at low $T$'s (the context we focus in this 
contribution).
We also discuss in full details the FEF protocol. 
In Sec. III we analyze the BEG, Bell-Lavis (BL), and 
Potts lattice models.
For the BEG and BL we also compare the FEF results with those for two 
other schemes (AP and CE) and illustrate the methods performance 
dependence on the number of replicas $R$.
Lastly, we present final remarks and conclusion in Sec. IV.

\section{The method details}

In general, for systems displaying first-order transitions at low 
temperatures or with a large jump in the order parameter 
\cite{books-first-order}, the distinct coexisting phases are separated by 
large free-energy barriers, exhibiting trapping and metastable states.
Hence, such cases are interesting instances to test the proposed scheme.
So, next we first give a brief account of the ST method and discuss an 
appropriate way to analyze strong first-order phase transitions.
Then, we pass to describe a FEF framework for the ST method.

\subsection{The Simulated Tempering (ST)}

The ST follows a twofold procedure.
First, at a certain $T_r$ (and during some established number of steps), 
a standard Metropolis prescription evolves a system of Hamiltonian 
${\cal H}$ throughout the phase space allowed microstates $\{\sigma\}$.
Second, an attempt for the change $T_{r'} \rightarrow
T_{r''}$ (with $r', r'' = 1, 2, \ldots, R$; $\beta_r = 1/(k_{B} \, T_r)$; 
and $\sigma$ the system state at the attempt time step) is drawn from
\begin{equation}
p_{r' \rightarrow r''} = 
\min \{ 1, \, \exp[(\beta_{r'} - \beta_{r''}){\cal H}(\sigma) 
+ (g_{r''} - g_{r'})] \}.
\label{pt-weight}
\end{equation}
This scheme is repeated a number $N$ of times.
Also, we consider only adjacent exchanges, i.e., $\Delta r = |r''-r'| = 1$.

According to Eq. (\ref{pt-weight}), the transition probability 
$p_{r' \rightarrow r''}$ strongly depends on the temperatures difference.
Larger $\beta_{r'} - \beta_{r''}$ leads to lower acceptance probabilities, 
whereas lower  $\beta_{r'} - \beta_{r''}$, although enhancing the exchanges,
may not be efficient since the generated configurations at $T_{r''}$  
in general will be similar to those at $T_{r'}$.
Therefore, conceivably there is a compromise between opposite
factors, implying in the existence of a best set ${\cal T}_R$.

Finally, we comment that in some ST implementations, the correct weights  
$g_r = \beta_r \, f_r$ (with $f_r$ the free energy) -- whose role is to 
assure an uniform visit to the distinct $T$'s -- are approximated 
\cite{pande,nguyen}.
For our examples, we obtain the $g$'s exactly by means of the approach 
in \cite{sauerwein,fiore3}. 
In short (full details in \cite{sauerwein,fiore2,fiore3,cluster2}), 
suppose a lattice model composed of $K$ layers of $L$ sites each. 
The total number of sites (or the volume) is then $V = L \times K$.
Also, assume the full Hamiltonian written in terms of these layers as
\begin{equation}                                                               
{\cal H} = \sum_{k=1}^K {\cal H}(S_k,S_{k+1}),                                  
\label{e14}                                                                    
\end{equation}
where $S_k \equiv (\sigma_{1,k},\sigma_{2,k}, \dots ,\sigma_{L,k})$ denotes
the $k$-th layer state configuration and $S_{K+1} = S_1$ (periodic boundary 
conditions).
The transfer matrix ${\bf T}$ is defined in such way that its elements 
are ${\bf T}(S_k, S_{k+1}) = \exp[-\beta {\cal H}(S_k,S_{k+1})]$.
Thus, in the thermodynamic limit (achieved already at relatively small
$V$'s \cite{fiore3}) $f_r = -\ln[\lambda^{(r)}]/(\beta_r L)$, with
\begin{equation}
\lambda^{(r)} = \left.\frac{\langle {\bf T}(S_k, S_{k+1} = S_k) \rangle}
{\langle \delta_{S_k, S_{k+1}} \rangle}\right|_{\beta = \beta_r}.
\end{equation}
In the above expression, the averages $\langle \ldots  \rangle$ are
direct calculated from usual MC simulations \cite{fiore3}.

\subsection{Characterizing the phase transition}

For strong first-order phase transitions -- our interest here -- usual 
order parameters (such as density and magnetization) and even other 
thermodynamic quantities like energy, are very well described by expressions 
in the following generic functional form \cite{fioreluzprl,fioreluzjcp2}
\begin{equation}                                                              
W \approx 
(b_1 + \sum_{n=2}^{\mathcal{N}} b_n \, \exp[-a_n y])/                
(1 + \sum_{n=2}^{\mathcal{N}} c_n \, \exp[-a_n y]).                  
\label{eq1}                                                                    
\end{equation}
For $\xi$ the control parameter, $y = \xi - \xi^{*}$ denotes the 
``distance'' to the coexistence point $\xi^{*}$ and $\mathcal{N}$ is the 
number of coexisting phases. 
The $c_n$'s are constants and the $b_n$'s are related to 
$\partial f_n/\partial \xi$, for $f_n$ the free energy of the coexisting 
phase $n$  \cite{fioreluzjcp2}.
Only the $a_n$'s are (linear) functions of $V$. 
As a remarkable consequence, by considering different system sizes,
all curves $W \times \xi$ cross at $\xi=\xi^{*}$.

Thus, if a numerical approach can properly simulate a given thermodynamic
quantity in a phase space region nearby the coexistence point, then 
simple fittings using Eq. (\ref{eq1}) -- for just few system sizes $L$'s -- 
can fully 
determine the phase transition thermodynamic properties (for many examples 
and a very detailed discussion about the usage of Eq. (\ref{eq1}), see 
Ref. \cite{fioreluzjcp2}).
We emphasizes that for first-order phase transitions at small $T$'s, the 
above expression is a rigorous result.
In this way, $W$ provides a very reliable test for the method.
Indeed, if the simulations lead to order parameters not having the shape 
in Eq. (\ref{eq1}), this is a strong indication of the algorithm 
inadequacy.

So, assume a fixed $L$ and $\xi$ around the phase transition point 
$\xi^{*}$.
We can verify the protocol convergence towards the order parameter steady 
value as well as the tunneling between the phases as the following.
Appropriate sampling of the relevant regions in the phase space is achieved 
when the $W$'s fluctuate mildly around $W_0$, with 
$\overline{W} \approx W_0$ (in our examples averaged over more than 100 
simulations runs). 
On the contrary, trapping in one of the $n$ phases (even for $\xi = \xi^{*}$)
will result in simulated $W$'s substantially differing from $W_0$, in fact
closer to the $W_n$ of phase $n$.
Another hint of a good performance is to have $W(L, \xi^{*}) \approx W_0(L)$ 
regardless of $L$.
Lastly, an extra checking is to calculate $\partial W/\partial\xi$
and $\chi = (\langle W^{2} \rangle - \langle W \rangle^2) \, V$.
Trapping implies in low fluctuations, i.e. $\chi \sim 0$, whereas 
frequent visit to the distinct regions gives 
$\chi^{*} = (\partial W/\partial \xi)|_{\xi^{*}}$. 
At the coexistence, $\chi^{*} \sim V$.  

\subsection{Obtaining an efficient set $\mathcal{T}_R$}

\begin{figure}
\setlength{\unitlength}{1.0cm}
\includegraphics[scale=0.33]{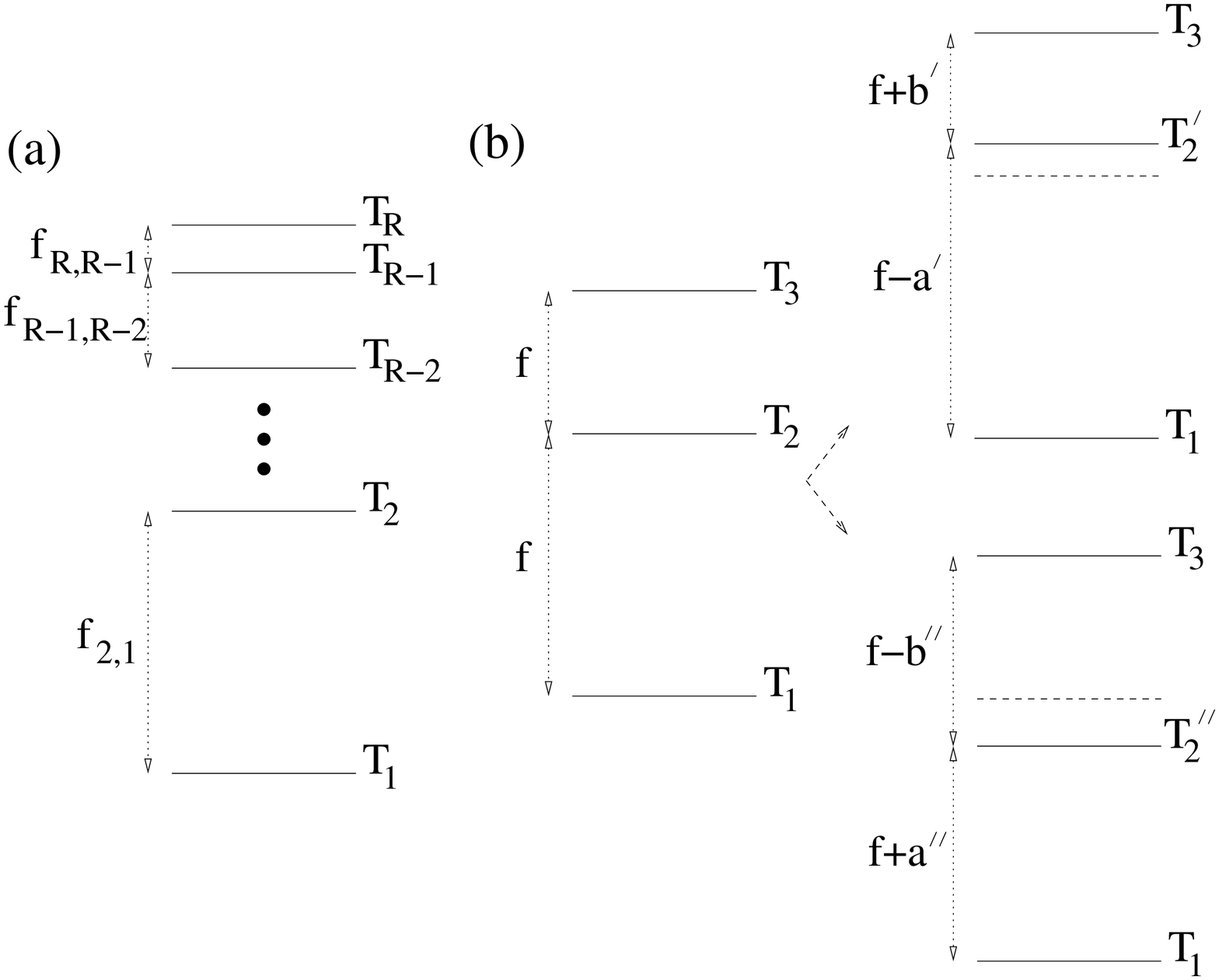}                
\caption {(a) Given ${\mathcal T}_R = \{T_1, \ldots, T_R\}$, successful 
attempts in changing the system temperature $T_r \rightarrow T_{r+1}$ will 
give rise to an exchange frequency $f_{r+1,r}$. 
The FEF protocol requires ${\mathcal T}_R$ to be chosen so that for 
any $r$, $f_{r+1,r} = f$.
(b) For $R=3$, small variations in the intermediate temperature $T_2$
will modify the FEF exchange frequency $f$ as indicated (with $a, b > 0$).}
\label{figure1}
\end{figure}

As already mentioned, given $R$, distinct key aspects -- specially for the 
ST \cite{note1} -- are involved in the $T_r$'s choice.
One example is the actual value of $T_R$.
Indeed, higher $T_R$'s facilitate the system to escape from the metastable 
states at $T_1$ (the temperature of interest).
But then, the exchange probability may be too low.
Conversely, the frequency of exchanges certainly increases for lower $T_R$'s, 
but this time the trapping might not be overcome.
Of course, one solution would be to use many replicas, however to the 
expense of lengthy simulations.

To simplify and improve the determination of ${\mathcal T}_R$, we propose 
the following protocol.
Once fixed $T_1$, we choose $T_1 < T_2 < T_3 < \ldots < T_R$ in such a 
way that the resulting exchange frequencies $f_{r+1, r}$ between any two 
successive temperatures $T_r$ and $T_{r+1}$  are all equal to some value 
$f$ (Fig. 1).
We define $f_{r+1, r} = N_{r+1,r}/N_{MC}$, with $N_{MC}$ the number of time steps 
in a Monte Carlo run.
Note that the highest temperature, $T_{R}$, is automatically 
established by the procedure, an advantage in contrast with some other 
recipes (see below).

The next step is to ``probe'' the efficiency of the obtained ${\mathcal T}_R$ 
by means of tests like those described in the end of Sec. II-B.
The existence of trapping indicates that $T_R$ is low and $f$ should be 
decreased, raising $T_R$. 
On the opposite, a very high $T_R$ (resulting from a too low $f$) implies 
in a very small fraction of exchanges and hence poor averages, so 
$f$ should be increased.
An optimal tuning (respect to the present recipe) will lead to
$f_{opt}$, given rise to a balanced tunneling between the phases and a faster
convergence to the steady state.
This would accelerate the calculations and improve the estimates
accuracy.

Finally, three technical aspects must be pointed out.
First,  some numerical work is necessary to find $f_{opt}$.
However, such process demands relative short simulations (e.g., it is not 
necessary to evolve the system until full convergence).
Thus, the search for the corresponding optimal ${\mathcal T}_R$ is not 
computationally time consuming.

Second, usually one shall study phase transitions in a region,
instead only at one point $\Lambda$, of the parameters space.
From the exploratory numerics we have performed (representative examples in 
Sec. III), we found that once an optimal set of temperatures 
is determined for a specific $\Lambda$, in a not too small vicinity around
it the same set also yields rapid and good results.
So, in exploring sectors of the parameters space one needs to find only few 
optimal ${\mathcal T}_R$'s.
In particular, ${\mathcal T}_R$ does not significantly change for 
reasonable different system sizes $L$.
Therefore, the procedure can be helpful for finite scale size analysis 
\cite{fioreluzprl}.

Third, $T_{R}$ for $f_{opt}$ is a value which at once allows enough temperatures
exchanges, yet resulting in proper tunneling between the phases.
Thus, an eventual further improvement of ${\mathcal T}_R$ should not 
drastically change $T_{R}$, just modifying the intermediate temperatures 
$T_2, \ldots, T_{R-1}$.
From our construction, at $f_{opt}$ the combined exchange frequency from 
$T_1$ to $T_R$ is roughly proportional to $F_{R,1} \sim (f_{opt})^{R-1}$.
Let us consider Fig. 1 (b) with $R=3$.
Increasing (decreasing) $T_2$, with $T_2 \rightarrow T_2'$ 
($T_2 \rightarrow T_2''$), the overall transition frequency 
$1 \rightarrow 3$ ($F_{3,1}^{(\Delta)}$) reads:
$F_{3,1}^{(\Delta)} \sim f_{opt}^2 + (b' - a') f_{opt} - b' a'$ 
($f_{opt}^2 + (a'' - b'') f_{opt} - b'' a''$).
Although we have not been able to derive expressions for the $a$'s and 
$b$'s, from exhaustive numerics we never found  $a$'s and $b$'s making 
$F_{1,3}^{(\Delta)}$ considerably larger than $F_{1,3}$. 
So, extra variations of the $T_r$'s around the values in the optimal 
${\mathcal T}_R$ makes the procedure much more cumbersome and does not
seem importantly to improve the method.

\section{Results}

Here we apply the FEF approach to three lattice models, BEG, BL, and 
Potts, also comparing some results with those for the ST using other 
schemes to select  ${\mathcal T}_R$. 
So, we need first to brief discuss few distinct methods to obtain the 
$T_r$'s.

Some proposals are rather simple, usually not taking explicit 
into account the physical aspect of the system (but refers to
\cite{predescu, kone} for the particular case of constant specific heats).
The exception being usually the numerical procedure to set the final 
temperature $T_R$ \cite{note2}.
Along this line, one example is to assume (given $R$, $T_1$ and $T_R$) the
intermediate temperatures forming an arithmetic progression (AP) 
\cite{calvo} (see also \cite{fiore2}).

A more thermodynamic-oriented method is the constant entropy (CE) protocol, 
proposed in \cite{sabo, kofke} (and extended in \cite{fiorejcp} for distinct 
lattice-gas models).
It consists of choosing intermediate $T_r$ values so to lead to a 
same fixed difference of entropy between successive temperatures
(but again, somehow $T_R$ must be first determined).
Specifically, if at $T_1$ and $T_{R}$ the system entropy per volume is, 
respectively, $s_1$ and $s_{R}$, we set $T_r$ such that 
$s_r = s_1 + (r-1) (s_R - s_1)/(R-1)$. 
Once the transfer matrix method of Sec. II-A gives the free energy, 
the evaluation of entropies for the CE protocol follows from 
$s_r = (u_r - f_r)/T_r$, where $u = \langle {\cal H} \rangle$.

Thus, an immediate advantage of FEF with respect to both AS and CE is 
that it directly provides an optimal $T_R$ necessary for the system to 
circumvent the entropic barriers. 
We also observe that close to the phase transition, where a small change 
of the control parameter results in a great variation of $s$, the FEF and 
the CE yield more clustered $T_r$'s.
Because quantities like $u$ and $s$ undergo pronounced changes around the 
phase transition, a clustered ${\mathcal T}_R$ is in general more 
efficient.
So, in advance one should expect CE and FEF better than AS.

\subsection{The models}
                               
\subsubsection{BEG}
\label{section-beg}

The BEG \cite{BEGMODEL} is a generalization of the Ising model, where 
the spin variable presents three values $\sigma_i = 0, \pm 1$. 
In a fluid language, it corresponds to a lattice-gas model with
distinct species ($\pm$) and vacancies ($0$).
The Hamiltonian also presents an extra interaction term, 
proportional to $\sigma_{i}^{2} \, \sigma_{j}^{2}$, reading
\begin{equation}                                                              
{\cal H} = - \sum_{<i,j>} 
(J \, \sigma_{i} \, \sigma_{j} + 
K \, \sigma_{i}^{2} \, \sigma_{j}^{2}) - 
\mu \, \sum_{i} \sigma_i^2,
\label{e3}                                                                     
\end{equation}
for $J$ and $K$ interaction energies and $\mu$ the chemical potential.
Its phase diagram is relatively complex, displaying phases with distinct 
structural properties. 
For the $K$'s we consider here and in the regime of low 
temperatures, a first-order transition separates liquid and gas phases 
for high and low chemical potentials, respectively.
The order parameter is the particle density 
$\rho = \langle \sigma_{i}^{2} \rangle$. 
At the steady state the system presents two liquid phases (with densities 
close to $1$) coexisting with one gas phase (of $\rho \approx 0$).
Since they have equal weights,  
$W_0 = \rho_0 = (1 \times 0 + 2 \times 1)/(1+2) = 2/3$. 
For $T = 0$, the liquid-gas phase coexistence emerges at 
$\mu^{*} = - z (K+1)/2$, with $z$ the coordination number. 
By increasing $T$, an order-disorder phase transition takes place, 
being continuous or discontinuous depending on how high is $T$. 

The transfer matrix, used to evaluate the weights for the ST method
and the entropy, is given by
\begin{eqnarray}                                                               
{\bf T}(S_{k}, S_{k+1} = S_k) &=& 
\exp\Big[\beta \sum_{i=1}^{L} \Big(J \, \sigma_{i+1,k} \, \sigma_{i,k}      
\nonumber \\                   
& &                                                
+ (J + \mu + K \, (1 + \sigma_{i+1,k}^{2})) \, \sigma_{i,k}^{2}                
\Big) \Big]. \nonumber \\                                                       
\label{eqsa}
\end{eqnarray}

\subsubsection{Bell-Lavis (BL)}

The BL is the simplest orientational model reproducing water-like 
features, including thermodynamics, dynamics and anomalous solubility 
\cite{fiore-m,fiore-m2,fiore-m3}.
It is defined on a triangular lattice and described by two kind of 
variables.
$\sigma_{i}$ determines if the site $i$ is either empty (0) or occupied
(1) by a molecule.
On the other hand, $\tau_{i}^{ij}$ indicates the possibility of hydrogen 
bonding formation between adjacent molecules.
Each molecule has six arms (in angles of 120$^{\circ}$), such that three 
of them are inert, while the other three are the bonding arms.
If at a site $i$ there is a molecule pointing its bonding arm towards the
site $j$, then $\tau_{i}^{ij}=1$, otherwise $\tau_{i}^{ij}=0$.
For $\epsilon_{vdw}$ and $\epsilon_{hb}$ the van der Waals and hydrogen
bonds interaction energies, the BL model is defined by the following 
Hamiltonian  
\begin{equation}                                                               
{\mathcal H} = - \sum_{<i,j>} \sigma_{i} \, \sigma_{j} \,                  
(\epsilon_{hb} \, \tau_{i}^{ij} \, \tau_{j}^{ji} +                    
\epsilon_{vdw}) - \mu \sum_{i} \sigma_{i}.                                     
\label{hambl}                                                                  
 \end{equation}

For $\zeta = \epsilon_{vdw}/\epsilon_{hb} > 1/3$ (the case we discuss here)
the system has three stable phases, gas, low-density-liquid (LDL) and
high-density-liquid (HDL) \cite{bell,fiore-m}, emerging as one increases 
$\mu$.
In the gas phase, molecules are scarce and unstructured ($\rho \approx 0$), 
whereas in the LDL phase (absent in the BEG model), the molecules are 
organized in a honeycomb network, with a density of $\rho=2/3$. 
In the HDL phase, the lattice is nearly fully filled with $\rho=1$. 
Also in contrast with the liquid phase for the BEG model, the HDL phase
is highly degenerated.
At $T = 0$, gas-LDL and LDL-HDL transitions are first-order and occur at 
$\mu^{*} = -3 (1 + \zeta)/2$ and $\mu^{*} = - 6 \zeta$, respectively.
For $T \neq 0$, the gas-LDL remains first-order, whereas LDL-HDL becomes 
second-order \cite{fiore-m}. 
For $\zeta=0.1$, the second- and first-order lines meet at a tricritical 
point.
The order parameter for the gas-LDL is the density, hence
$W=\rho$.
At the phase coexistence, $\rho_0=1/2$ (understood from the fact that the 
LDL phase has degeneracy $3$ and since their weights become equal at the 
phase coexistence, the value 
$\rho_0 = (1 \times 0 + 3 \times 2/3)/(1+3) = 1/2$ follows). 

The transfer matrix reads
\begin{eqnarray}                                                               
{\bf T}(S_{k},S_{k+1} = S_k) &=&
\exp\Big[\beta \sum_{i=1}^{L} \Big(                                             
\sigma_{i,k} \, (\sigma_{i,k} + 2 \sigma_{i+1,k}) 
\nonumber \\
& & \times                          
(\epsilon_{vdw} + \epsilon_{hb} \, \tau_{i,k} \,                              
\tau_{i+1,k} + \mu) \big) \Big].
\end{eqnarray}

\subsubsection{Potts model}

The Potts \cite{pottsreview} is a simple spin lattice model for which 
the variable $\sigma_i$ (defined on the site $i$) takes the integer 
values $0, 1, \ldots, q-1$. 
Adjacent sites $i'$ and $i''$ have a non-null interaction energy of 
$-J$ whenever $\sigma_{i'} = \sigma_{i''}$.
The problem full Hamiltonian reads
\begin{equation}
{\cal H} = -J \sum_{<i,j>} \delta_{\sigma_i  \sigma_j}.
\end{equation}

For small temperatures the system is ordered, becoming disordered as
$T$ increases. 
The transition point is exactly given by
$T_{c} = 1/\ln[1+\sqrt{q}]$. 
In 2D, for $q \leq 4$ the phase transition is of second-order and 
discontinuous if $q \geq 5$. 
A proper order parameter $W = \phi$ is 
\begin{equation}
\phi = \frac{q (V_{max} / V)-1}{q-1},
\label{order-potts}
\end{equation}
where $V_{max}$ is the volume occupied by the spins of the state $s$ of
largest population and $V = L^2$.

For the Potts, the transfer matrix yields
\begin{equation}                                                               
{\bf T}(S_{k},S_{k+1} = S_k)=
\exp\Big[\beta J\sum_{i=1}^{L} \Big(1 + 
\delta_{\sigma_{i+1,k}  \, \sigma_{i,k}} \Big) \Big].    
\end{equation}

\subsection{Simulations}

For the numerics we consider regular lattices with periodic boundary 
conditions, also setting the Boltzmann constant $k$ equal to 1.
For the BEG and Potts we assume a square and for the BL a triangular 
lattice of volume $V = L \times N$.
The parameters used are:
$T_1=0.5000$, $K/J = 3$, $J=1$ (BEG),
$T_1=0.1000$, $\zeta = 0.1$, $\epsilon_{hb} = 1$ (BL), and
$q=20$, $T_1 = 1/\ln[1+\sqrt{20}] = 0.5883\ldots$, $J=1$ (Potts).
Moreover, unless otherwise explicit mentioned $L=N=20$, $L=N=18$
and $L=N=18$ for, respectively, the BEG, BL and Potts models.
In all cases we are in the low temperature regime and for BEG and BL
strong discontinuous transitions (with coexisting phases separated 
by large free energy barriers) result from a chemical potential 
$\mu_0$ close to its $T=0$ value, so that  
${\mu}_0=-8.0000(1)$ at $T_1=0.5000$ (BEG) and 
${\mu}_0=-1.6500(1)$ at $T_1=0.1000$ (BL). 
Finally, we recall that for larger $T$'s, obviously the phase coexistence 
takes place for distinct values of the chemical potentials
(e.g., for the BL and $T=0.25$, the gas-LDL phase transitions is at 
$\mu = -1.6528(1)$).
In those cases the proposed protocol also works better than other
procedures to choose the replicas temperatures (as we have explicit 
verified).
But since then the entropic barriers are lower, the present very strong 
first-order phase transition is a more interesting context for our 
comparative analysis.

\subsubsection{Results using FEF for the BEG and BL models}
\label{sec-iii-b-1}

For the BEG model, in Tables \ref{table1} and \ref{table2} we show 
for $R=3$ and $R=4$ some sets ${\mathcal T}_R$ obtained with the 
FEF protocol.

\begin{table}
\begin{tabular}{cccccc}
f & $5 \times 10^{-2}$ & $1 \times 10^{-2}$ & $6 \times 10^{-4}$ ($f_{opt}$) 
& $10^{-5}$  \\
\hline
$T_1$ & 0.50 & 0.50 & 0.50 & 0.50    \\
$T_2$ & 1.35 & 1.45 & 1.60 & 1.82    \\
$T_3$ & 1.70 & 1.88 & 2.05 & 2.33    \\
\end{tabular}
\caption{For the BEG model, temperature sets ${\mathcal T}_{R=3}$ for 
distinct frequencies $f$.\label{table1}}
\end{table}

\begin{table}
\begin{tabular}{ccccc}
f  & $1.1 \times 10^{-1}$ & $5 \times 10^{-2}$ 
& $2 \times 10^{-2}$ $(f_{opt})$ & $10^{-4}$  \\
\hline
$T_1$ & 0.50 & 0.50 & 0.50 & 0.50  \\
$T_2$  & 1.25 & 1.35 & 1.40 & 1.45  \\
$T_3$ & 1.55 & 1.70 & 1.80 & 1.88  \\
$T_4$ & 1.78 & 1.95 & 2.04 & 3.20  \\
\end{tabular}
\caption{The same as in Table \ref{table1}, but for $R=4$.\label{table2}}
\end{table}

\begin{figure}
\setlength{\unitlength}{1.0cm}
\includegraphics[scale=0.35]{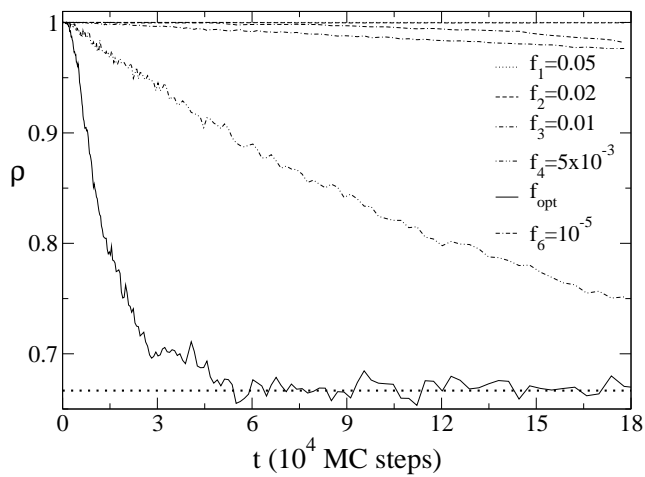}                
\caption {For the BEG model at the phase coexistence (see main text), 
$\rho$ simulated with the ST algorithm for the ${\mathcal T}_{R=3}$'s 
obtained from distinct $f$'s.
The best convergence toward the steady value $\rho_0 = 2/3$ is 
obtained for $f_{opt} = 6 \times 10^{-4}$.}
\label{fig1aa}
\end{figure}

To determine which set ${\mathcal T}_{R}$ is the best one, in Fig. 
\ref{fig1aa} we compare for the BEG model and $R=3$ the density evolution
towards the steady value $\rho_0 = 2/3$ (we start from
a non-typical initial configuration, a lattice totally filled of 
particles).
For larger $f$'s (e.g., $f_1$, $f_2$ and $f_3$), despite more frequent 
temperature exchanges, the system gets trapped in the initial configuration 
as a consequence of a too low $T_3$.
On the other hand, for much lowers $f$'s (as $f_6$), the resulting 
$T_3$'s become high enough to cross the entropic barrier, but then 
exchanges hardly take place. 
Hence, there is an optimal intermediate value, $f_{opt}$, yielding the
best convergence to the correct $\rho_0$. 

We also shall observe the order of magnitude of the time scale in Fig. 
\ref{fig1aa}, i.e., of $10^{4}$ MC steps.
Note in the graph that for $f_{opt}$, the very fast decay of $\rho$ 
to its correct value is already evident for small values of $t$.
Therefore, even simulations for times around $10^{3}$ MC would be able 
to determine $f_{opt}$.
However, the apparently longer than necessary simulation in Fig.
\ref{fig1aa} has a practical reason.
As discussed in \cite{fiore2}, specially for the ST method, the 
computation time $\overline{t}$ need to estimate thermodynamical 
quantities generally must be longer (at least one order of magnitude) 
than $\tau$.
Here, $\tau$ is the typical time to overcome the transients and to reach 
the steady state (e.g., in Fig. \ref{fig1aa}, 
$\tau \approx (3$--$7) \times 10^{4}$). 
In fact, because the many temperatures changes during the whole simulation, 
the ST requires a proper $\overline{t}$ to ensure enough sampling 
at the desired $T_1$. 
This point, directly related to tunneling between the phases and exchange 
rates between the $T_r$'s, becomes even more relevant at low $T$'s.
Thus, Fig. \ref{fig1aa} is also helpful to give an idea about 
$\overline{t}$.
For instance, for the BEG model, we have $\overline{t} = 3 \times 10^{5}$ 
MC steps.

\begin{figure}[h]
\setlength{\unitlength}{1.0cm}
\includegraphics[scale=0.35]{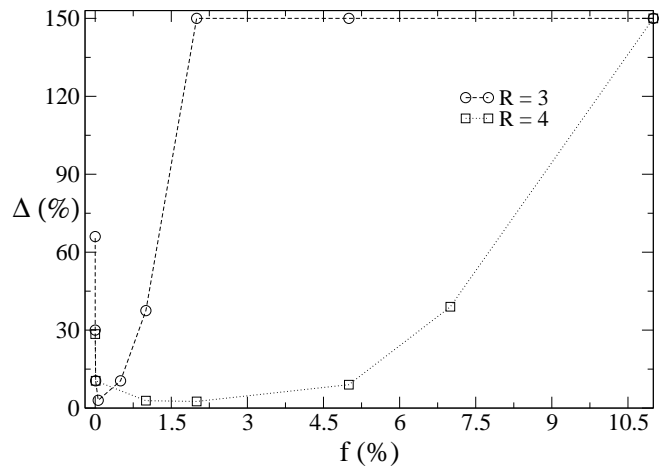}                
\caption{$\Delta = |2/3 - \langle \rho \rangle|/(2/3) \times 100 \%$
as function of $f (\%)$. 
$\Delta$ is a minimum, of $3\%$ and $2.5\%$, respectively for $R=3$
and $R=4$, when $f = f_{opt}$.}
\label{figure3}
\end{figure}

Using the above $\overline{t}$ to make the averages 
$\langle \rho \rangle$ for the simulated $\rho$, in Fig. \ref{figure3} 
we plot for the BEG model and  $R=3$ and $R=4$, the percentile difference 
from $\rho_0 = 2/3$, 
$\Delta = |2/3 - \langle \rho \rangle|/(2/3) \times 100 \%$, as function 
of $f (\%)$.
As it should be, $\Delta$ has a minimum for $f = f_{opt}$.
We also see that $f_{opt}$ is greater for $R=4$.
This is a direct consequence of the $T_{R}$'s to be essentially the same 
when $f=f_{opt}$ (compare Tables \ref{table1} and \ref{table2}).
So, $T_{r+1} - T_{r}$ is smaller for $R=4$, yielding a higher $f_{opt}$.
At the optimal condition, $\langle \rho \rangle$ is equal to
0.665(2) and 0.666(1) for $R=3$ and $R=4$, respectively. 

Similar analysis have been done for the BL model.
In Tables \ref{table3} ($R = 3$) and \ref{table4} ($R = 4$) we 
illustrate the results showing few values of $f$ and the associated 
${\mathcal T}_R$'s.
From plots like those in Figs. \ref{fig1aa} and \ref{figure3} 
(not shown here) we have determined the $f_{opt}$'s as indicated in the
tables.
Furthermore, for $f_{opt}$ we have estimate that 
$\overline{t} = 5 \times 10^6$ MC steps.
This longer time necessary for the averages reflects a higher complexity 
of the BL model, hence more difficult to simulate than the BEG.
At the optimal condition, $\langle \rho \rangle = 0.508(5)$ 
($R=3$) and $\langle \rho \rangle = 0.500(4)$ ($R=4$).

\begin{table}[h]
\begin{tabular}{cccccc}
f & $2.5 \times 10^{-2}$ & $3 \times 10^{-3}$ & $2 \times 10^{-4}$
($f_{opt}$)
& $10^{-5}$  \\
\hline
$T_1$ & 0.10 & 0.10 & 0.10 & 0.10    \\
$T_2$ & 0.25 & 0.28 & 0.32 & 0.35   \\
$T_3$ & 0.33 & 0.38 & 0.43 & 0.48    \\
\end{tabular}
\caption{For the BL model, temperature sets ${\mathcal T}_{R=3}$ for
distinct frequencies $f$.\label{table3}}
\end{table}

\begin{table}[h]
\begin{tabular}{ccccc}
f  & $1.5 \times 10^{-1}$ & $7 \times 10^{-2}$
& 1$ \times 10^{-2}$ $(f_{opt})$ & $10^{-5}$  \\
\hline
$T_1$ & 0.10 & 0.10 & 0.100 & 0.10  \\
$T_2$  &0.20 & 0.23 & 0.27 & 0.29  \\
$T_3$ & 0.25 & 0.30 & 0.34 & 0.39 \\
$T_4$ & 0.29 & 0.39 & 0.43 & 0.50  \\
\end{tabular}
\caption{The same as in Table \ref{table3}, but for $R=4$.\label{table4}}
\end{table}

\begin{figure}
\setlength{\unitlength}{1.0cm}
\includegraphics[scale=0.32]{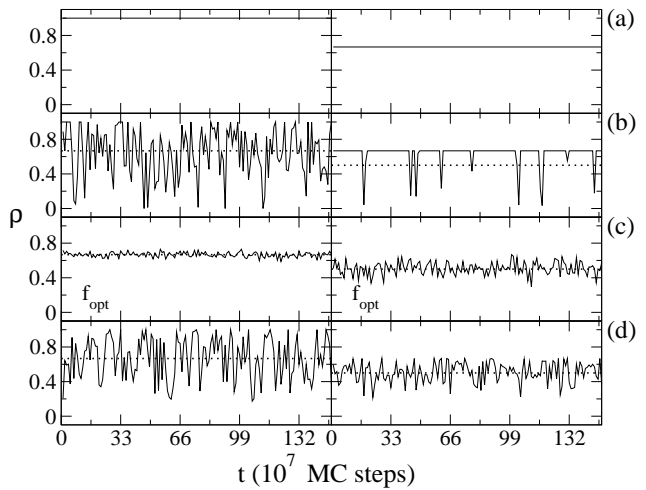}                
\caption {For $R=3$, $\rho$ versus the simulation time for $t$ much latter 
than $\overline{t}$. 
The results are for the BEG (left panel) and BL (right panel) models,
where (a), (b), (c), and (d) represent the four sets ${\mathcal T}_R$
in the same order presented in Tables \ref{table1} (BEG) and 
\ref{table3} (BL).}
\label{fig1}
\end{figure}

\begin{figure}
\setlength{\unitlength}{1.0cm}
\includegraphics[scale=0.32]{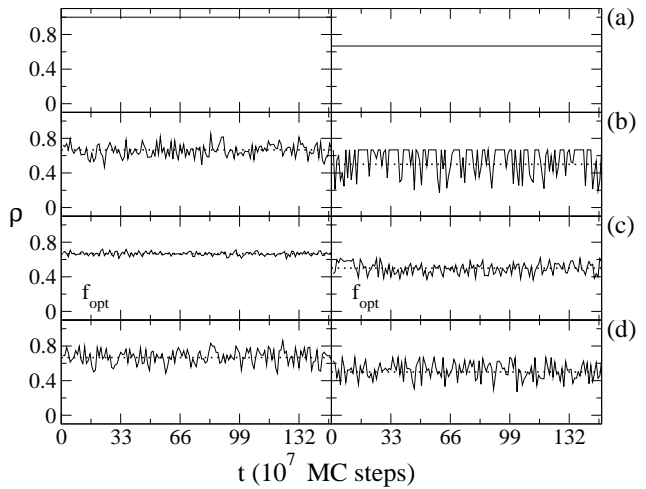}                
\caption {The same as in Fig. \ref{fig1}, but for $R=4$.
(a), (b), (c), and (d) are for the sets in Tables \ref{table2} (BEG) 
and \ref{table4} (BL).}
\label{fig2}
\end{figure}

Another important factor to guarantee proper estimates for the thermodynamic 
quantities is to assure a frequent tunneling between the phases {\em after} 
the system to reach the steady state.
By plotting $\rho$ for times much larger than $\overline{t}$, we can further 
check when this condition is indeed fulfilled for the difference choices 
of $f$, Tables \ref{table1}--\ref{table4}.
The simulations are shown in Figs. \ref{fig1} and Fig. \ref{fig2},
$R=3$ and $R=4$, with the left (right) panels displaying the BEG (BL) 
model.
For the largest $f$'s, (case (a) in both figures) the sets ${\mathcal T}_R$ 
are rather inefficient and the system stays trapped into one phase.
For the $f$'s in (b) and (d), the density substantially fluctuates about
its true equilibrium value $\rho_0$. 
It is exactly for $f_{opt}$ that $\rho$ oscillates much closer to $\rho_0$ 
(case (c)), thus leading to the best estimates.
As one should expect, overall the better results are those for a 
larger $R$, compare Figs. \ref{fig1} and \ref{fig2}.
This is a consequence of a more balanced sampling due to a larger number
of replicas, but then requiring greater computation efforts.

\begin{figure}
\setlength{\unitlength}{1.0cm}
\includegraphics[scale=0.35]{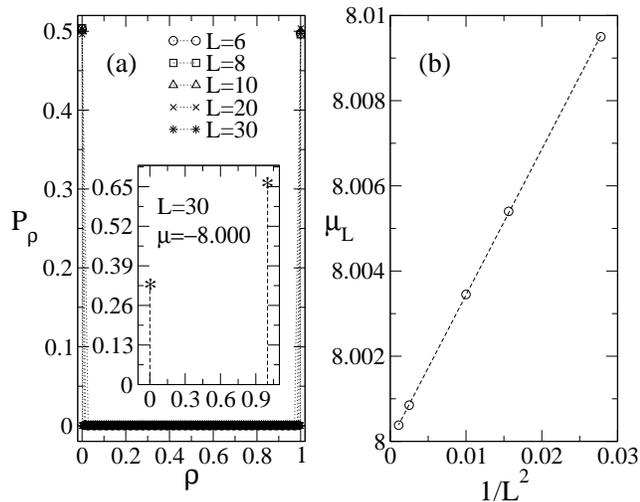}                
\caption {For the BEG model, (a) the probability distribution histogram
$P_\rho$ of the order parameter $\rho$ at the coexistence for distinct 
system sizes $L$.
The different distributions -- each calculated at a given chemical 
potential value $\mu_L$ -- are clearly bimodal (with the gas phase having
$\rho \approx 0$ and the two liquid phases having $\rho = 1$).
In the inset, $P_\rho$ for $\mu = -8.000$ and $L=30$.
In (b), the $\mu_L$'s in (a) vs. the inverse of the volume $1/L^2$.
Extrapolating to the thermodynamic limit (i.e., infinite volume), one gets
$\mu = -8.00000(5)$, in very good agreement with $\mu_0 = -8$ 
(main text).}
\label{rep1}
\end{figure}

\begin{figure}
\setlength{\unitlength}{1.0cm}
\includegraphics[scale=0.35]{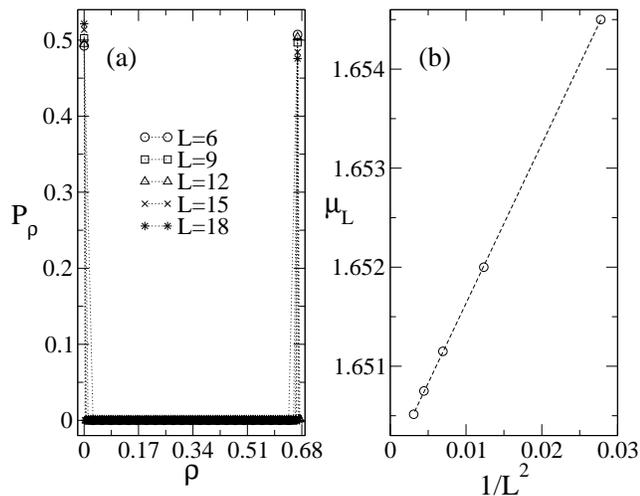}                
\caption {The same as in Fig. \ref{rep1}, but for the BL model.
Here, an extrapolation from (b) leads to $\mu = -1.65000(5)$, 
in agreement with $\mu_0 = -1.65$.}
\label{rep2}
\end{figure}

Lastly, to verify that at the transition condition (of first-order) in 
fact the system is properly visiting the coexisting phases, we perform a 
final important test.
After the convergence to the steady state (using $R=3$ and the optimal set 
$\mathcal{T}_3$), we calculate the probability distribution for $\rho$ at 
the temperature of interest $T_1$, assuming different system sizes $L$.
For each $L$, by properly setting $\mu = \mu_L$ we have the results
for the BEG and BL models, respectively, shown in Figs. \ref{rep1} and 
\ref{rep2}.
In all cases, we see that the probability distributions $P_\rho$ display 
well defined bimodal shapes, with the peaks at the $\rho$ values of the 
individual phases.
This clearly indicates a first-order phase transition.
Moreover, from simple scaling arguments \cite{books-first-order}
(typically used to locate the coexistence points, see, e.g., 
\cite{wang}), the chemical potential values $\mu_L$ should vary linearly 
with the inverse of the volume (i.e., $\mu_L - \mu_0 \sim 1/L^2$) and 
give the correct $\mu_0$ in the thermodynamic limit.
This is indeed observed in Figs. \ref{rep1} (b) and \ref{rep2} (b), whose
extrapolated $\mu$ for $1/L^2 \rightarrow 0$ agrees very well with the 
analysis using the function $W$ in the end of Sec. III-B-2.

An interesting aspect, which can be analyzed from the probability 
distribution plots, is how the combined distinct phases $\rho$'s, 
at the coexistence, lead to the system steady state density $\rho_0$.
Let us consider the BEG model as an example.
As already discussed in Sec. III-A-1, for the BEG $\rho_0 \approx 2/3$ 
since exactly at transition we have two phases with $\rho=1$ and one with 
$\rho \approx 0$, all contributing with a same weight.
This is indeed verified in Fig. \ref{fig1aa}. 
In Fig. \ref{rep1} (a), for different $\mu_L$'s and $L$'s, the $P_\rho$ is 
strongly concentrated in $\rho = 0$ and $\rho = 1$, as it should be.
However, the peaks heights are not in the proportion 1:2.
This is so because for the $\mu$'s and $L$'s considered in the graphs, 
we are not still at the thermodynamic limit, when the equal probability 
for the phases does hold.
In contrast, for $\mu = -8.000$, 
$P_\rho$ in the inset of Fig. \ref{rep1} (a) exhibits the expected 1:2 ratio.
Although not explicit shown, this behavior is also the case for the BL
model in Fig. \ref{rep2}.

Finally, we observe that as discussed in details in Refs. 
\cite{fiore2,fiore3} for the BEG and in Refs. \cite{fiore3,fioreluzjcp2} 
for the BL, these systems also go through strong first order phase 
transitions for temperatures three times as higher as the ones used here. 
Nevertheless, in such cases the choice of ${\mathcal T}_R$ is not so 
critical as for the $T$'s in the present work (see, e.g., the discussion 
in \cite{fiorejcp}).

\subsubsection{Comparison with other ${\mathcal T}_R$ schemes for the 
BEG and BL}

Now, we compare the FEF with some other available protocols to set 
${\mathcal T}_R$, namely, AP and CE.
To facilitate their implementation (recall that AP and CE do not have a 
specific rule to determine the maximum temperature), we use for them the 
same value of $T_R$ found from the FEF in the optimal condition.

\begin{figure}
\setlength{\unitlength}{1.0cm}
\includegraphics[scale=0.32]{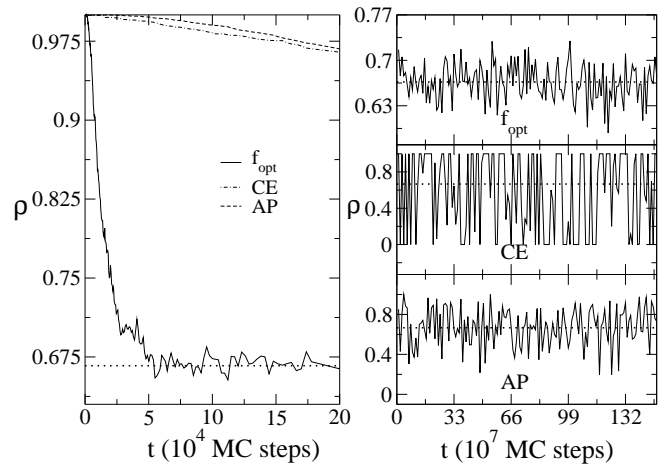} 
\caption{For the BEG model, simulations from the ST method with $R=3$ 
replicas and different schemes to determine ${\mathcal T}_R$.
Left panel: $\rho$ during the transient time (towards the steady state).
Right panel: $\rho$ versus time (in MC steps) -- already at the steady 
state -- showing how each temperature set makes the system to tunnel 
between the coexisting phases (observe the different $\rho$ scales in 
each plot).
The dotted lines indicate the correct $\rho_0=2/3$.
The temperature set for AP follows directly (since $T_1$ and $T_{R=3}$ are
given) and the intermediate temperature for the CE is $T_2 = 1.88$.
Also, $\langle \rho \rangle_{CE} = 0.61(5)$ and 
$\langle \rho \rangle_{AP} = 0.67(2)$.}
\label{fig5-b}
\end{figure}

\begin{figure}
\setlength{\unitlength}{1.0cm}
\includegraphics[scale=0.32]{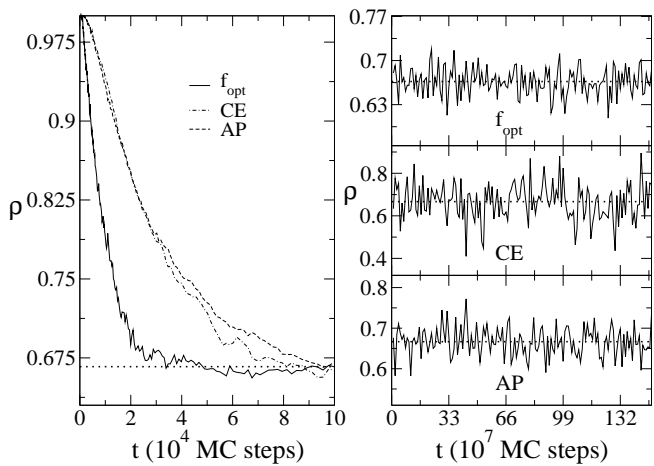} 
\caption{The same as in Fig. \ref{fig5-b}, but for $R=4$.
For the CE scheme, the intermediate temperatures are $T_2 = 1.75$ and 
$T_3 = 1.95$.
Also, $\langle \rho \rangle_{CE} = 0.661(6)$ and 
$\langle \rho \rangle_{AP} = 0.666(3)$.}
\label{fig5-c}
\end{figure}

For the BEG model (for the BL the results are similar, thus not shown here) 
we analyze the density time evolution toward the steady state and the 
tunneling between coexisting phases in Figs. \ref{fig5-b} ($R=3$) and 
\ref{fig5-c} ($R=4$).
It is clear that the FEF recipe provides the fastest convergence towards 
$\rho_0$. 
In addition, the tunneling between the phases is also substantially more 
frequent than for AP and CE.

\begin{figure}
\setlength{\unitlength}{1.0cm}
\includegraphics[scale=0.275]{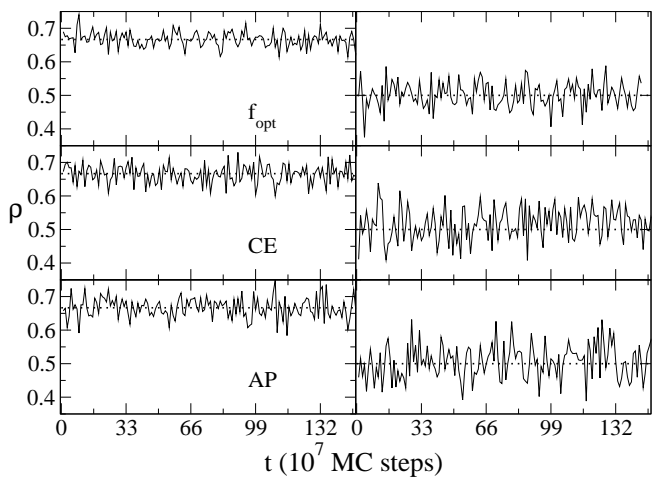} 
\caption{$\rho$ versus $t$ for the BEG and BL, left and right panels,
at phase coexistence and distinct ${\mathcal T}_R$ schemes, with
$R=6$.
In all cases the extreme temperatures are $T_1 = 0.5$, $T_6 = 2.06$ (BEG)
and $T_1 = 0.1$, $T_6 = 0.43$ (BL).
For the $f_{opt}$ and CE recipes, the intermediate temperatures are,
respectively: 
$\{1.25, 1.55, 1.78, 1.95\}$ and $\{1.61, 1.82, 1.94, 2.00\}$ (BEG);
and
$\{ 0.23, 0.30, 0.32, 0.39 \}$ and $\{ 0.31, 0.36, 0.40, 0.41 \}$ (BL).}
\label{fig6-c}
\end{figure}

Although the FEF is systematically better than AP and CE, 
the difference is less pronounced for $R=4$.
This should be expected since for a larger number of replicas, 
provided $T_R$ is properly chosen, the exact values of the intermediate
$T_r$'s do not play a so critical role.
To further illustrate this fact, we repeat the simulations of Figs.
\ref{fig5-b} and \ref{fig5-c}, but now using $R=6$.
This is displayed in Fig. \ref{fig6-c} for BEG and BL models, where it
is shown the tunneling frequency after the transient.
Visually, the distinct schemes seems to give the same results
(but note a smaller statistical uncertainty, fluctuation, for the FEF).
We get $\langle \rho \rangle$ from the $f_{opt}$, CE and AP,
respectively equal to, 0.667(2), 0.666(2), and 0.666(2) for the BEG and 
0.501(3), 0.52(1), and 0.51(1) for the BL.
These values should be compared to $\rho_0 = 2/3 = 0.666\ldots$ for the 
BEG and $\rho_0 = 0.5$ for the BL.
Actually, the FEF still gives a little more accurate estimates, specially 
for BL.

\begin{figure}
\setlength{\unitlength}{1.0cm}
\includegraphics[scale=0.28]{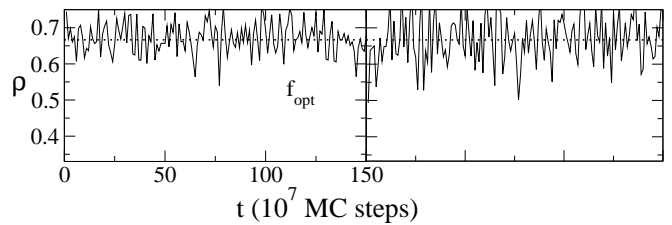} 
\caption{$\rho$ versus $t$ for the BEG model at phase coexistence for 
the $R=6$ optimal temperature set and two large $L$ values.}
\label{fig7al}
\end{figure}

Although one of the nice features of the `combo' approach here --
to combine Eq. (\ref{eq1}) with the ST using an optimal choice for
${\mathcal T}_R$ -- is to be able to obtain the thermodynamic limit
for discontinuous transitions from considerably small systems 
\cite{fioreluzprl,fioreluzjcp2} (see next), the same scheme in principle
does also work well for large $L$'s.
For instance, for the BEG model, we consider in Fig. \ref{fig7al} the
previous optimal set obtained for $R=6$ replicas (Fig. \ref{fig6-c}),
but now simulating sizes of $L=30$ and $L=40$. 
As it can be seen, the proper tunneling between coexisting phases 
confirms that a same optimal ${\mathcal T}_R$ can be used for 
distinct big $L$'s.

\begin{figure}
\setlength{\unitlength}{1.0cm}
\includegraphics[scale=0.3]{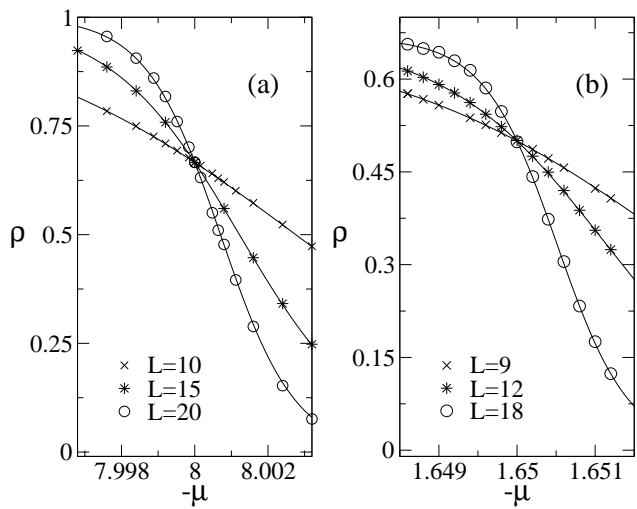} 
\caption{$\rho$ versus the chemical potential $\mu$ (for distinct
$L$'s) for the (a) BEG and (b) BL models.
The simulations (symbols) are performed with the ST method, for $R=3$
and the set ${\mathcal T}_R$ corresponding to $f_{opt}$ in 
Tables \ref{table1} and \ref{table3}.
The continuous lines are fits using the general functional form $W$ for 
$\rho$ (where only four simulated points are necessary to determine
the coefficients $a$, $b$ and $c$ in Eq. (\ref{eq1})).}
\label{fig7}
\end{figure}

As already stated, some initial numerical work necessary to determine 
the optimal ${\mathcal T}_R$ (for a specific point $\Gamma$ of the
parameters space) pays off not only because then the full simulations 
are faster and more reliable, but also because this same set could be 
used for other points in the vicinity of $\Gamma$.
To exemplify this, we consider the ST method with $R=3$ and use the 
optimal ${\mathcal T}_{R=3}$ in Tables \ref{table1} (BEG) and 
\ref{table3} (BL) -- determined for $\mu = \mu_0$ and $L$ as in Sec. 
\ref{sec-iii-b-1}.
Then, we calculate $\rho$ for distinct values of the chemical potential 
$\mu$ and system size $L$.
The results are shown as symbols in Fig. \ref{fig7}.
In all the simulations we have obtained fast and good convergence to the 
steady state and frequent tunneling between phases exactly as in our 
previous analysis.
The continuous curves are fits given by the general expression $W$, 
Eq. (\ref{eq1}), which is the correct shape of the order parameter $\rho$ 
in a first-order phase transition at low temperatures \cite{fioreluzprl}.
The very good matching between the simulated points and the smooth curves 
$W$ illustrate that indeed $\rho$ versus $\mu$ for different $L$'s is 
well described by our approach from just one ${\mathcal T}_R$.
Furthermore, the crossings take place at $\mu = -8.000(1)$ (BEG) and
$\mu = -1.6500(1)$ (BL), as it should be (see Sec. III-B-1).

\subsubsection{The FEF method for the Potts model}
\label{sec-iii-b-3}

\begin{table}[h]
\begin{tabular}{ccccc}
f  & $4 \times 10^{-1} $ & $1 \times 10^{-1}$
& 5$ \times 10^{-2}$ $(f_{opt})$ & $2 \times 10^{-2}$  \\
\hline
$T_1$ & 0.5885 & 0.5885 & 0.5885 & 0.5885  \\
$T_2$  &0.5945 & 0.6030 & 0.6054 & 0.6100  \\
$T_3$ & 0.5972 & 0.6245 & 0.6260 & 0.6300 \\
\end{tabular}
\caption{For the Potts model, temperature sets ${\mathcal T}_{R=3}$ for
distinct frequencies $f$.\label{table5}}
\end{table}

The Potts is an interesting (and also challenging) system to analyze 
given the existence of strong first-order phase transitions for high 
values of its parameter $q$.
Hence, the model is frequently used as a benchmark to test different 
numerical algorithms (see, for instance, Refs. 
\cite{neuhaus-magiera,kim-straub-keyes,kim-straub,fioreluzjcp2}).
So, next we present some of our previous discussions regarding the 
efficiency of the FEF protocol considering the Potts with $q=20$ 
(therefore, already a large value in the above mentioned respect).

\begin{figure}
\setlength{\unitlength}{1.0cm}
\includegraphics[scale=0.33]{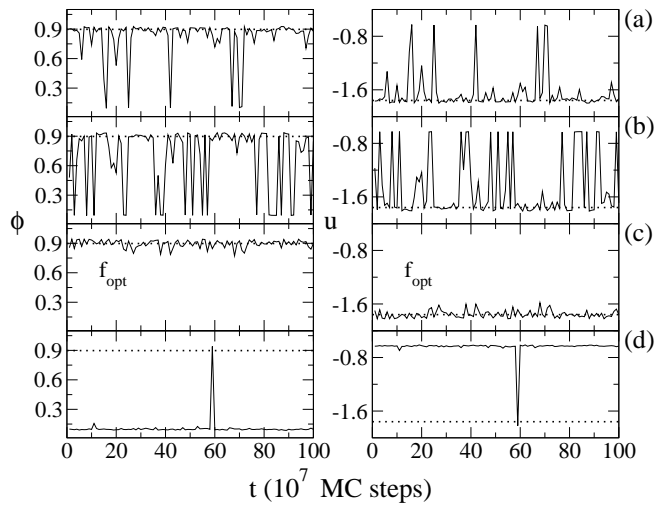}                
\caption {For $R=3$, $\phi$ and $u$ versus the simulation time $t$ 
for the Potts model. 
The results in (a), (b), (c), and (d) represent the four sets 
of ${\mathcal T}_{R=3}$ displayed in Table \ref{table5}.}
\label{fig12}
\end{figure}

For $R=3$, in Table \ref{table5} we show the temperatures values for 
distinct choices of $f$.
Due to the strong transition presented by the problem for $q=20$, from the 
simulations determining the $f$'s we estimate $\overline{t} = 10^7$ MC 
steps, thus higher than, e.g., $\overline{t} \sim 10^5$ obtained for the 
BEG model.
In Fig. \ref{fig12} we plot the order parameter $\phi$, Eq. 
(\ref{order-potts}), and the energy $u = \langle {\cal H}\rangle$ as 
function of $t$ for distinct $f$'s.
Averages are evaluated each ${\bar t=10^{7}}$ MC steps. 
Dotted lines are steady values obtained from a very precise approach
(and more complex than the present one) developed in \cite{fioreluzjcp2}, 
which combines cluster algorithms, the PT method and semi-analytic protocols
to enhance tunneling between the different coexisting phases.

\begin{figure}
\setlength{\unitlength}{1.0cm}
\includegraphics[scale=0.35]{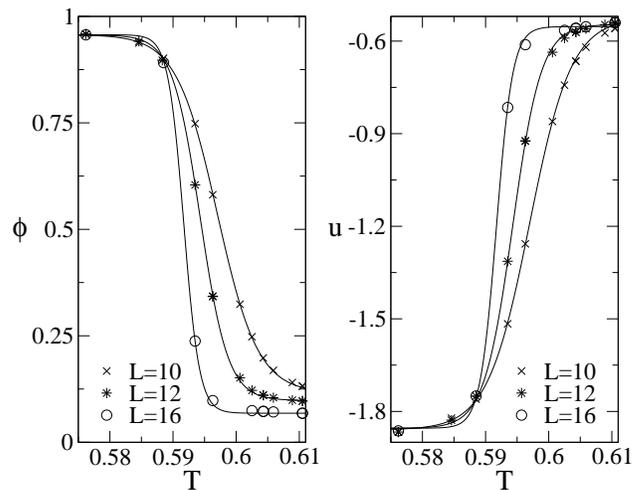}                
\caption {For the Potts model and $R=3$, $\phi$ and $u$ versus the 
temperature $T$ for three values of $L$.
The simulations (symbols) are performed with the set ${\mathcal T}_R$ 
corresponding to $f_{opt}$ in Table \ref{table5}.
The continuous lines are fits using the general functional form $W$
and only four simulated points.}
\label{fig13}
\end{figure}

The behavior of both $\phi$ and $u$ in Fig. \ref{fig12} is in agreement
with that observed for the BEG and BL order parameters. 
There is an optimal value of $f$, $f_{opt}$, allowing substantially more 
frequent tunnelling between phases and thus much better statistics to
calculate relevant thermodynamic quantities. 
On the other hand, ``fortuitous'' choices for ${\mathcal T}_R$ (arbitrary
$f$'s) yield poor results and an eventual impression that the usual ST 
approach could not handle strong first order phase transitions.

Lastly, we perform an analysis similar to that shown in Fig. \ref{fig7} for 
the BEG and BL models.
For Fig. \ref{fig13} we: consider distinct $T$'s around the $q=20$
Potts transition temperature, use the same $f_{opt}$ in Table \ref{table5} 
-- found for $T=0.5883\ldots$ -- to form sets ${\mathcal T}_R=3$ (with 
$T_1 = T$), and calculate $\phi$ and $u$ as function of such $T$'s.
Again we see that the present method provides very precise values (as it 
can be confirmed, e.g, by comparing with Ref. \cite{fioreluzjcp2}) for 
both order parameter and energy (well described by Eq. (\ref{eq1})). 
However, we should mention that in \cite{fioreluzjcp2} it was 
necessary to combine cluster algorithms and the PT (with non-adjacent 
exchanges and larger number of replicas) for the system to properly cross
the coexisting phases.
Hence, the results here have been obtained with much less computational
efforts.

It is also worth contrasting ours with some studies combining 
generalized ensembles with the PT \cite{kim-keyes-straub} and ST 
\cite{kim-straub} algorithms.
In these works, the methods efficiency for the Potts model are
indeed very good, nevertheless comparable with the ST-FEF in terms of 
numerical precision and computational time.
Also, in Refs. \cite{kim-keyes-straub,kim-straub} $q=8$, hence a weaker 
transition than here.
However, the simulations are done for systems up to $L = 64$, 
so much larger than the present ones ($L = 18$).
But they use $R=30$ replicas. 
In Figs. \ref{fig12} and \ref{fig13}, we have been able to obtain proper 
convergence just with $R=3$ and small systems (the latter in part 
thanks to the use of our semi-analytic fitting, Eq. (\ref{eq1})).

\section{Final remarks and conclusion}

In this contribution we have proposed a simple method to determine the set 
of $T_r$'s for the ST algorithm.
For a given number of replicas $R$, adjacent $T_r$'s are chosen in such a 
way that changes of temperature (from $T_r$ to $T_{r+1}$, for any 
$r=1,2,\ldots,R-1$) occur with a fixed frequency $f$.
Then, from simulations demanding a relative small number of time steps
we test, e.g., the  decay toward the steady regime and the tunneling rate 
between distinct coexisting phases.
Such tests can benchmark the efficiency of ${\mathcal T}_R$ resulting from 
this $f$.
Repeating the procedure, it is possible to find an optimal $f_{opt}$ 
leading to a best ${\mathcal T}_R$ (including a proper $T_R$).
Note that for some other schemes, there is no specific rules to 
determine the extreme temperature necessary for the system to cross the 
entropic barriers.

The searching for $f_{opt}$ demands some preliminary numerics, nevertheless 
twofold compensated.
First, the full simulations -- for the final estimation of the 
thermodynamical quantities -- will converge faster and with greater 
accuracy.
Second, the same optimal ${\mathcal T}_R$ can be used to simulate
the system for other parameters values, around the original $\Gamma$
used to determine $f_{opt}$.

The reliability and precision of the FEF scheme have been analyzed in the 
extreme situation of systems presenting first-order phase transitions 
at low $T$'s.
In all examples we have obtained very good thermodynamic estimates 
(even for $R=3$) without the necessity of long simulations.
In fact, one of the advantages of using the FEF to select
${\mathcal T}_R$ is that once at the optimum case, the results are already 
very good for $R$ small, hence facilitating the numerical implementation.
Furthermore, comparison with other temperature schemes, AP and CE, have 
been undertaken. 
The FEF has been found to be always superior to the others, specially 
for small $R$'s.

To conclude, we shall comment on the following.
First, in the last few years distinct protocols have been proposed for 
dealing with the difficult situation of first-order transitions at low $T$'s.
In particular, Ref. \cite{neuhaus-hager} shows how to increase visits to 
regions of high free-energies (hence with very low probability to be 
accessed) considering additional multiple Gaussian weights for the ensembles 
describing the system.
Moreover, a combination with the parallel tempering is used for restoring 
ergodicity.
In Ref. \cite{kim-straub} it is devised an adaptive (``on the flight'') 
approach for the ST weights, whose evaluations follow a modified distribution 
\cite{tsallis}.
In fact, using the ST with independent modified-weight runs, the tunneling 
between the phases is strongly facilitated.

Although certainly the above methods are of very broad usage and important 
contributions to handle such hard problem, our proposal somehow strikes 
in a different direction. 
Actually, not only the free energy weights are evaluated directly from 
standard MC simulations (so, without no effective resampling procedures, 
which eventually could depend on systems particularities), but also adequate
``probings'' of the phase space distinct regions are achieved solely 
from the tempering.
As shown, the crucial point is to determine a good temperature set, 
possible through relative inexpensive simulations.
Thus, our approach is easier to implement and can be faced as a 
generic algorithm, not requiring too much details about the specific system.
The mentioned methods are very efficient, but may demand elaborate 
implementations focusing the specific case at hands (e.g., properly
tuning dynamic dependent parameters). 

Second, although we have used an optimized protocol to determine 
${\mathcal T}_R$ for a tempering method, applying it to first-order 
transitions (where the specific $T_r$'s may be crucial), we believe the 
main ideas here can also be useful for second-order phase transitions.
Indeed, in second-order phase transitions the trapping in metastable states 
is not generally present.
However, at the criticality the system is strongly affected by a slow 
time decay of correlation functions (critical slowing down)
\cite{fiore2}. 
In such case, an optimal ${\mathcal T}_R$ might provide a faster decay
of correlations, thus allowing a more precise evaluation of the
system properties. 
In fact, a practical way to achieve so would be to find $f_{opt}$ 
which minimizes the relaxation time $\tau$ of a given relevant 
correlation function.

Third, in the so called parallel tempering (PT), usually exchanges
of temperature are not restricted to adjacent replicas ($T_r$ and $T_{r+1}$),
hence the exact set ${\mathcal T}_R$ is not so crucial
(but see \cite{rathore-2005}).
Nevertheless, a simpler algorithm using only adjacent exchanges eventually
would require a more appropriate choice for ${\mathcal T}_R$, which could 
then be determined by the FEF.

The above two possible applications for our method are presently being
implemented and will be reported in the due course.

\section{Acknowledgment} 
We acknowledge research grants from CNPq and computational facilities
provided by CT-Infra-Finep and CNPq-Edital Universal.
  


\begin{thebibliography}{99}
\bibitem{note1}
For a given ${\mathcal T}_R$, contrary to the PT, in the ST just to 
consider non-adjacent exchanges does not necessary improve the protocol 
efficiency.
\bibitem{note2}
One way to define $T_R$, when no other procedure is available,
is to apply simple Metropolis algorithms to successive temperatures
above $T_1$, picking the lowest one $T$ for which the method is capable 
to give satisfactory results for that $T$.
\end{thebibliography}
\end{document}